\documentclass[journal,onecolumn,draftclsnofoot,]{IEEEtran}
\usepackage{ifpdf}

\usepackage{cite}

\usepackage{graphicx}

\usepackage{amsmath}
\usepackage{amssymb}
\usepackage{subfig}

\usepackage{bm}
\usepackage{fixltx2e}
\usepackage{tikz}

\newcommand\copyrighttext{%
	\footnotesize \textcopyright 2020, Elsevier. Licensed under the Creative Commons Attribution-NonCommercial-NoDerivatives 4.0 International http://creativecommons.org/licenses/by-nc-nd/4.0/ 
\begin{center}
	\includegraphics[width=0.15\linewidth]{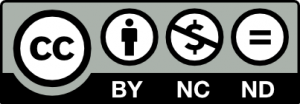}
\end{center}

}
\newcommand\copyrightnotice{%
	\begin{tikzpicture}[remember picture,overlay]
	\node[anchor=south,yshift=10pt] at (current page.south) {\fbox{\parbox{\dimexpr\textwidth-\fboxsep-\fboxrule\relax}{\copyrighttext}}};
	\end{tikzpicture}%
}

\begin{document}
\title{Parametric Amplification of Broadband Vibrational Energy Harvesters for Energy-Autonomous Sensors Enabled by Field-Induced Striction}
\author{{Ulrike Nabholz, Lukas Lamprecht, Jan E. Mehner, Andr\'{e} Zimmermann and Peter Degenfeld-Schonburg}
\thanks{U. Nabholz, L. Lamprecht and P. Degenfeld-Schonburg are with Robert Bosch GmbH, Corporate Research, 71272 Renningen, Germany (e-mail: ulrike.nabholz@de.bosch.com).}
\thanks{J. E. Mehner is with Chemnitz University of Technology, 09107 Chemnitz, Germany.}%
\thanks{A. Zimmermann is with University of Stuttgart and Hahn-Schickard, 70569 Stuttgart, Germany.}%
}
\maketitle
\begin{abstract}
We investigate the influence of parametric excitations on MEMS vibration energy harvesters for energy autonomous sensor systems. In Industry 4.0 (or Industrial IoT) applications, interconnected sensors provide a means of data acquisition for automated control of the manufacturing process. Ensuring a continuous energy supply to the sensors is essential for their reliable operation. Manufacturing machines usually display a wide spectrum of vibration frequencies which needs to be covered by an array of harvester substructures in order to maintain the desired output level. 
We show that mechanical structures designed to implement a Helmholtz-Duffing oscillator have an increased bandwidth by exploiting several orders of parametric resonances. In contrast to concepts implementing parametric amplification in a multi-mode scenario, our concept is based on a single mechanical mode. Therefore, it is more robust against fabrication tolerances as the relevant multi-mode resonance conditions do not need to be matched on the level of single chips. 
Using exact transient simulations and semi-analytic models to showcase the relation of the Helmholtz-Duffing oscillator to the damped and driven Mathieu equation, we show that parametric resonances highly increase the bandwidth of the output power whenever high Helmholtz nonlinearities are present. 
To achieve the required nonlinearities, we suggest nonlinear stress-strain curves and we propose to achieve such nonlinearities through field-induced striction by magneto- or electrostriction. In contrast to existing approaches, where external fields are harvested using strictive effects, we employ external fields that manipulate the effective Young's modulus to achieve parametric excitations in a mechanical oscillator. Thus, we are able to propose a novel energy harvester concept incorporating strictive materials that exploits the effects of parametric excitations to achieve broadband vibrational energy harvesting. 
\end{abstract}
\copyrightnotice\\

\section*{Highlights}
\noindent
1.) Parametrically amplified harvesting of ambient field energy for wireless sensor hub\\
2.) Parametric excitations in a single mode Helmholtz-Duffing oscillator\\
3.) Nonlinear stress-strain curves achieved through field-induced striction\\
4.) Increased achievable broadband power of vibrational energy harvester
\IEEEpeerreviewmaketitle
\section{Introduction}
\IEEEPARstart{E}{nergy-efficient}
wireless sensor systems are proposed as a key element for digitalization in the Internet of Things (IoT) \cite{Farhan2019}. Within the last decade, research on IoT has been booming \cite{Liu2017} and application areas like smart home or smart manufacturing are promising future markets for IoT hardware like wireless sensor nodes (WSN) \cite{Xu2018, Lu2017}. Predicted market volumes for Industry 4.0 (I4.0) (or Industrial IoT) related sensors and actuators of \$4.6 billion -- a plus of 70\% compared to 2017 -- indicate the importance of maintenance-free energy supplies for WSN \cite{Yole2018}. To avoid cost intensive and often hardly realisable cabling as well as battery replacements, energy harvesting is a commonly proposed technique \cite{Shaikh2016, Akhtar2015}. This work presents a new approach for broadband kinetic energy harvesting to maximize energetic lifetime with parametric amplification using stricitve material to harvest vibrational energy boosted by magnetic or electric fields. This concepts aims at industrial environments like electrical drives or transformers that typically offer both types of ambient energy. In the following, we introduce the basic concept of kinetic energy harvesting and briefly analyse competing methods for broadband and nonlinear harvesting methods. We introduce the concept of parametric amplification in more detail throughout the work.
\subsection{Kinetic Energy Harvesting for Industrial Applications}
\noindent Industrial kinetic energy harvesting can use rotational or translational \cite{Lamprecht2018} as well as vibrational energy. Vibration energy harvesters (VEH) typically use relative motions between seismic masses and vibrating peripheral components that are connected via transducer elements. Most common are piezoelectric, inductive or capacitive transducers, while other approaches like triboelectric harvesters have less relevance due to their lack of maturity \cite{Lamprecht2019}. IoT- and I4.0-related applications of energy autonomous sensor systems powered by VEHs are often affected adversely by changes in environmental conditions. For resonant based harvester approaches, matching the dominant excitation frequency of the IoT-device with that of the VEH leads to maximized power outputs due to direct harmonic resonance. Therefore, VEHs often comprise a single resonance frequency at which they can be operated that is set by their design. Contrarily, industrial applications like motors or pumps often change their operating points -- e.g. due to changing shaft speeds or variable mechanical loads. Since acceleration peaks in vibration spectra are strongly correlated to shaft speeds, fixed resonance VEHs are not reasonable for IoT and I4.0 application. Instead, broadband and frequency tuning approaches are investigated in both academia and industry \cite{Zhu2010, Daqaq2014, Lamprecht2019}. Besides complex designs and required internal controls for actively frequency tuned VEHs, energy consumption is a main disadvantage of resonance-following approaches. On the other hand, passive strategies for achieving broadband harvesting
often lead to voluminous designs. Array configurations with separate substructures for fixed frequencies lead to reduced power densities. Using multimodal approaches leads to rather less gain in harvested energy due to lower displacements at higher frequencies \cite{Heshmati2019}. A more lucrative way
for achieving broad operational bandwidths is to use nonlinearities \cite{Abdelkefi2012}.
\subsection{Parametric Amplification}
\label{pa}
\noindent Parametric amplification (PA) uses nonlinear effects that cause parametric resonances besides the linear direct resonances and therefore allows to harvest vibrational energy at higher displacements \cite{Caldwell2017}. 
A major challenge for designing parametrically amplified VEHs is to include large frequency ranges within which a higher magnitude of the nonlinear oscillation can be reliably achieved than in the linear case. 
Yet, Duffing-like harvesters in stiffer configurations for higher frequencies can hardly take advantage of nonlinearity reliably. This is due to the fact that the new state of a Duffing-type oscillator depends on the initial conditions of the system which usually cannot be predicted in energy harvesting applications: In the region of the amplitude response curve close to resonance, two stable solution branches exist for a single frequency during steady-state. As detailed in \cite{Daqaq2014}, whether the system response converges to the upper or lower solution branch depends on the initial conditions. \\
For energy harvesting of wireless sensor systems, the amplitude and rate of change in the amplitude of the harvested vibrations is variable and thus the initial conditions at each point in time are hard to predict. Assuming the worst case for energy harvesting, the exploitable displacement would not exceed the linear case.  \\
Cantilever-based VEHs using parametric resonances can be realised with a clamped-clamped configuration \cite{Yildirim2016}, as well as clamped-free configurations \cite{Yildirim2017, Daqaq2009} in longitudinal direction. For linearly operated cantilever VEHs, typically transversal operation is desired due to highest deflection and therefore highest output power e.g. using piezoelectric transducers. Further enhancements regarding broadband operation can be achieved by multi-modal approaches. 
Both macroscopic designs \cite{Jia2013a} and MEMS-based parametrically amplified VEHs still have to overcome initiation threshold amplitudes of parametric resonance \cite{Jia2013b}. These will be called 'critical points'.
This paper investigates how parametric resonances can be used with simple single-clamped beams or membranes in transversal operation using a single mode approach with the aim of achieving low values for the critical points of parametric resonance.
\subsection{Design approach}
\label{design}
\noindent In contrast to existing literature \cite{Mori2015, Davino2016, Clemente2017, Deng2017}, where striction is used to harvest external magnetic or electric fields, we use external fields in order to tune a mechanical oscillation by manipulating the effective Young's modulus and drive our harvester design in the regime of parametric amplification as will be detailed in section \ref{theory}.
\begin{figure}[h]
	\centering
	\includegraphics[width=0.7\linewidth]{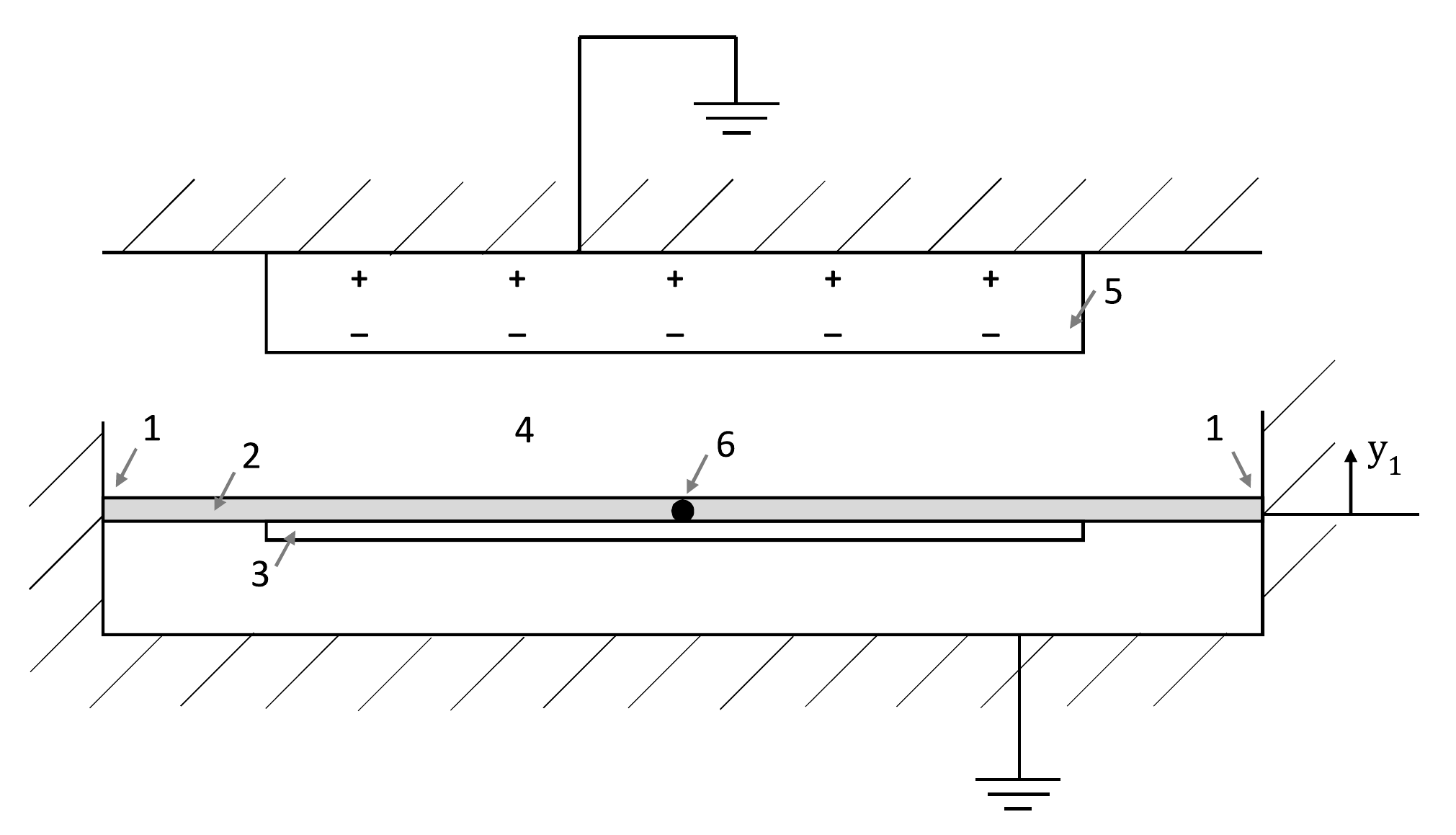}
	\caption{Generalized design approach for a parametrically amplified VEH using field-induced striction.}
	\label{fig:design}
\end{figure}
Fig. \ref{fig:design} shows a possible operating principle of a parametrically amplified VEH using field-induced striction. The key elements of the design are denoted by numbers: (1) shows the a clamped support as one possible support method. Other support methods are feasible, in particular pivot and sliding support. In principle, different designs can be constructed using all possible combinations of the three support elements on either side of the structure. (2) denotes the structure oscillating in \(y_\mathrm{1}\)-direction (called out-of-plane direction) to which the VEH (3) is attached. Note that the strictive effects are only applied to the strictive layer (2) and that no electrostrictive effects occur for the piezoelectric transducer elements conventionally used in VEHs. Structure (5) represents the electrode coated in electret, a dielectric material with static electric charge as indicated.\\
The actuation can be implemented in several ways, the black circle (6) denotes the point of application of the force acting on the oscillating structure. The force originates from the external vibration which we aim to harvest and is depicted as a total resulting force. The direction of force can either be parallel to the out-of-plane direction \(y_\mathrm{1}\) or perpendicular to it. The sense of direction can be constant (corresponding to a static force) or alternating. \\
Our generalized design approach provides us with a large range of possible combinations of support methods (e.g. clamped-clamped or clamped-free), geometry of the oscillating structure (e.g. beams or membranes), material type (e.g. metallic or elastomeric) and force application (e.g. sinusoidal or shock). These possess individual advantages depending on the application and the constraints concerning cost, space and power requirements. \\
Different properties of the design influence strictive effects and thus also the effective Young's modulus of the system (as Section \ref{struct} will show): Mainly, varying the amount of strictive material (e.g. the thickness of the strictive layer) influences the strength of the resulting striction forces. Furthermore, different strictive materials exhibit differently strong strictive properties. In general, increasing the thickness of the structures (i.e. membrane- or beam-thickness) increases the resonance frequency of the oscillator. 
\section{Theory}
\label{theory}
\noindent In this section, we describe the nonlinear dynamics behind our proposed single-mode parametrically amplified VEH principle. 
We find that the mode does not only oscillate in accordance with an external vibration frequency, but if certain conditions are met, also with an additional frequency close to the mode’s resonance frequency. The oscillation amplitude of the additional frequency is parametrically amplified, thus increasing the performance of the energy harvester \cite{Nitzan2015}, \cite{Jia2016}.\\
First, we introduce the structural mechanics that describe our system using a nonlinear strain energy formulation in Section \ref{struct}. Then, in Section \ref{theory_dimless}, the Helmholtz-Duffing oscillator is introduced as our model for a parametrically amplified VEH and the stability conditions are derived in Section \ref{stability}. Sections \ref{parexc}-\ref{subcrit} focus on establishing an analytical model based on Floquet-theory of the parametric excitations in our system, as well as on deriving representations of the critical and sub-critical behaviour. 
%
%
%
\subsection{Structural mechanics for harvesting nonlinearities}
\label{struct}
\noindent We described in Section \ref{pa}, how a parametrically amplified VEH for broadband harvesting applications designed as simple Duffing-type oscillator can hardly take advantage of its inherent nonlinearity. As measured by \cite{Jia2016}, material nonlinearities induced by doped silicon can yield several orders of parametric resonance. In Section \ref{design}, we proposed field-induced striction as an alternative design approach for reliably achieving the desired level of nonlinearity. \\
As we will introduce in Section \ref{theory_dimless}, our design is governed by equation \eqref{eq:duff_eom}. In order to achieve broadband harvesting, a system with a high asymmetric Duffing, or rather Helmholtz, coefficient is needed. It is possible to increase the Helmholtz coefficient through a nonlinear stress-strain relation. Here, we derive how a nonlinear stress-strain relationship provides the main handle for tuning our harvester system and exploiting the phenomenon of parametric excitations. 
Based on the derivation of the strain energy we carried out in \cite{Nabholz2019a}, where a linear stress-strain relation with the Piola-Kirchhoff stress field \(\bm{S}_{nm} = \bm{D}_{nmlk} \bm{E}_{lk}\) with the Green-Lagrange strain field \(\bm{E}_{lk}\) \cite{Kim2014, Sathyamoorthy1997} and the constant fourth-order material tensor \(\bm{D}_{nmlk}\) was assumed, we expand our model to allow for a nonlinear stress-strain relation, where a sixth order material tensor denoted by \(\bm{\tilde{D}}_{nmlkl'k'}\) appears:
\begin{equation}
\label{eq:stress_field}
\bm{S}_{nm} = \bm{D}_{nmlk}  \bm{E}_{lk} + \bm{\tilde{D}}_{nmlkl'k'} \bm{E}_{lk} \bm{E}_{l'k'} + \mathcal{O}\left(\bm{E}^3\right).
\end{equation}
with \(\mathcal{O}\left(\bm{E}^3\right)\) denoting the higher order terms which will be neglected. The Green-Lagrange strain field is given by
\begin{equation}
\bm{E}_{lk} = \frac{1}{2}\left(\underbrace{\frac{\partial u_l}{\partial y_k} + \frac{\partial u_k}{\partial y_l}}_{=\epsilon_{lk}} + \frac{\partial u_n}{\partial y_l} \frac{\partial u_n}{\partial y_k}\right),
\end{equation}
where \(\epsilon_{lk}\) is the linear component of the strain tensor. \(u_i\) denotes the i-th component of the displacement \(\bm{u}\), and \(y_i\) the i-th component of the material point \(\bm{y}\).\\ 
The cubic component, a nonlinear strain energy term \(\bm{U}_{\textrm{strain}}^{(3)}\) (cubic in the displacement) can be formulated based on equation \eqref{eq:stress_field} as 
\begin{equation}
\bm{U}_{\textrm{strain}}^{(3)} = \frac{1}{2} \left(\int_{V_\textrm{0}} 2 \bm{\epsilon}_{nm} \bm{D}_{nmlk} \bm{E}_{lk} + \int_{V_\textrm{0}}\bm{\epsilon}_{nm} \bm{\tilde{D}}_{nmlkl'k'} \bm{\epsilon}_{lk} \bm{\epsilon}_{l'k'}\right),
\end{equation}
where the first and second addends denote the strain energy due to geometric and material nonlinearities, respectively. \\
\(V_\textrm{0}\) denotes the volume of the undeformed structure, since the Piola-Kirchhoff stress field denotes the stress in relation to the undeformed domain.
The conversion between modal and nodal basis can be written by relating the (nodal) displacement and the sum of the modal terms \cite{Chopra1996}:
\begin{equation}
u_i\left(\bm{y}, t\right) = \sum_{n=1}^{N} \left(f_n\right)_i x_n\left(t\right) \: \text{for } i = 1, 2, 3
\label{eq:modal}
\end{equation}
with \(\left(f_n\right)_i\), the i-th component of the natural modes \(\bm{f}_n\), as well the modal coordinates \(x_n\left(t\right)\) in a system with \(N\) degrees of freedom.\\
For clarity of the explanations, we will only consider a single spacial dimension from here on. This simplification is justified, since our proposed model is only actuated in one direction whilst undergoing negligible deformations in all other directions. Restating equation \eqref{eq:stress_field} under this condition leads to the one-dimensional stress \(\sigma_\textrm{s}\) defined as 
\begin{equation}
\label{eq:youngsmod}
\sigma_\textrm{s}\left(\epsilon\right) = Y_0 \epsilon + Y_1 \epsilon^2 + 2_1 \epsilon^3= \underbrace{\left( Y_0 + Y_1 \epsilon + Y_2 \epsilon^2 \right)}_{= Y\left(\epsilon\right)} \epsilon.
\end{equation}
The second order Taylor approximation of the Young's modulus \(Y\) as a function of the linear strain \(\epsilon\) for nonlinear behaviour has been assumed by e.g. \cite{Yang2016, Kaajakari2004},
with the constant part of the Young's modulus \(Y_0\), first order term \(Y_1\), second order term \(Y_2\) and the corresponding strain \(\epsilon\). Higher order terms are neglected. In our modelling approach, the second order Young's modulus term \(Y_2 \epsilon^2\) leads to the same polynomial form as the cubic Duffing nonlinearity in the equation of motion that will be introduced in Section \ref{theory_dimless}. Thus, these nonlinear dynamics are already captured by the geometric nonlinearity of the Duffing term. \\
On the other hand, the component of \(Y_1\) in equation \eqref{eq:youngsmod} is directly proportional to the asymmetric Duffing coefficient, i.e. the main nonlinear tuning parameter in our model, see section \ref{theory_dimless}. In symmetrical structures, this term does not originate from geometric nonlinearities and can only result from material nonlinearities. Consequently, strategies to increase \(Y_1\) which entails increasing \(\epsilon\) are in the centre of interest. Proposed methods include doping of bulk silicon \cite{Yang2016} or, as we suggest here, field-induced striction. \\
Field-induced striction, such as magnetostriction or electrostriction, describes deformation due to an external magnetic or electric field. Magnetostriction is a property of all ferromagnetic materials, whereas electrostriction occurs for all ferroelectric materials. \\
Depending on the combination of magnetic or electric field strength and engineering stress in the mechanical structure, the Young's modulus of strictive materials varies. For more details, the reader is referred to \cite{Datta2010}, Fig. 4.
\subsection{Mathematical model of a parametrically amplified single-mode VEH}
\label{theory_dimless}
\noindent In this section, we will introduce the physical model to clarify our idea of a single-mode parametrically amplified vibrational energy harvester. The equation of motion is given by the Helmholtz-Duffing equation \cite{Kovacic2011}
\begin{equation}
\label{eq:duff_eom}
\ddot{x} + \omega_{\mathrm{0}}^{2}x + \frac{\omega_{\mathrm{0}}}{Q}\dot{x} + \alpha x^2 + \beta x^{3}  = g \cdot \sin\left(\omega_{\mathrm{exc}} t\right)
\end{equation}
with the angular resonance frequency \(\omega_0\), Quality factor \(Q\), Duffing coefficient \(\beta\), external acceleration amplitude \(g\) and angular excitation frequency \(\omega_{\mathrm{exc}}\). Without loss of generality we have set the arbitrary phase of the vibrational acceleration to zero. For the modal deflection that was introduced in equation \eqref{eq:modal} the simplified notation \(x\) is used from now on. Yet, the deflection is still time-dependent. In addition to the standard Duffing oscillator term \(\beta x^3\), we include an asymmetric force term \(\alpha x^2\) with the Helmholtz coefficient \(\alpha\). Equation \eqref{eq:duff_eom} has been formulated such that the physical units of the coefficients \(\alpha, \beta\) are given by \(\left[\alpha\right] = \mathrm{m^{-1} s^{-2}}\) and \(\left[\beta\right] = \mathrm{m^{-2} s^{-2}}\). 
Due to the simplicity of our one degree-of-freedom model we will use full transient-time simulations which accurately reproduce the physics of this model. However, it is useful to supplement these transient simulations by approximate analytical results obtained using different simulation strategies, namely steady-state, critical, and sub-critical approximations. These will be derived in the following sections and their results will be compared in Section \ref{simulation} for certain points in the parameter space. Our insights support the understanding and therefore the control of the major ingredients needed for a successful design conception of a parametrically amplified VEH.
In order to obtain analytical insights, it proves useful to proceed using dimensionless parameters \cite{Strogatz2007} which allows us to compare individual terms quantitatively. By introducing the dimensionless displacement \(z = x \frac{\omega_{\mathrm{0}}^2}{g}\) and dimensionless time \(\tau = \omega_{\mathrm{0}}t\), we can restate equation \eqref{eq:duff_eom} by 
\begin{equation}
\label{eq:duff_eom_dimless}
z'' + \frac{1}{Q} z' + z + \sigma z^2 + \frac{\sigma^2}{4} \eta z^3 = \sin\left(r \tau \right)
\end{equation}
with the dimensionless frequency ratio 
\begin{equation}
r = \frac{\omega_{\mathrm{exc}}}{\omega_{\mathrm{0}}}, 
\label{eq:ratio_r}
\end{equation}
the acceleration ratio 
\begin{equation}
\sigma = \frac{g \alpha}{\omega_{\mathrm{0}}^4} \label{eq:sigma}
\end{equation}
and the nonlinearity ratio 
\begin{equation}
\eta = \frac{4 \beta \omega_{\mathrm{0}}^2}{\alpha^2}. \label{eq:eta}
\end{equation} 
Note that the time-derivatives in equation \eqref{eq:duff_eom} and equation \eqref{eq:duff_eom_dimless}, indicated by dot and prime symbol, are applied with respect to the time \(t\) and the dimensionless time \(\tau\), respectively. \\
Most importantly, the power transferred to the VEH is defined by \(P = \frac{\omega_{\mathrm{0}}}{Q} m \dot{x}^2 \) \cite{Zhu2011}, with \(m\) denoting the mass of the VEH. Consequently, the time-averaged power is given by \(\overline{P} = \frac{\omega_{\mathrm{0}}}{Q} m \overline{\dot{x}^2} \). In order to assess the performance of the  harvester system, we define a dimensionless performance measure \(p\) as
\begin{equation}
\label{eq:performance}
p = \frac{1}{Q}\left(z'\right)^2,
\end{equation}
which relates to the time-averaged power as \(p = \frac{\overline{P}}{P_0}\) with \(P_0 = \frac{m g^2}{\omega_{\mathrm{0}}}\), \([P_0] = \mathrm{\frac{J}{s}}\).\\
\subsection{Stability conditions for the oscillatory motion}
\label{stability}
\begin{figure}[htbp]
	\centering
	\includegraphics[width=0.5\linewidth]{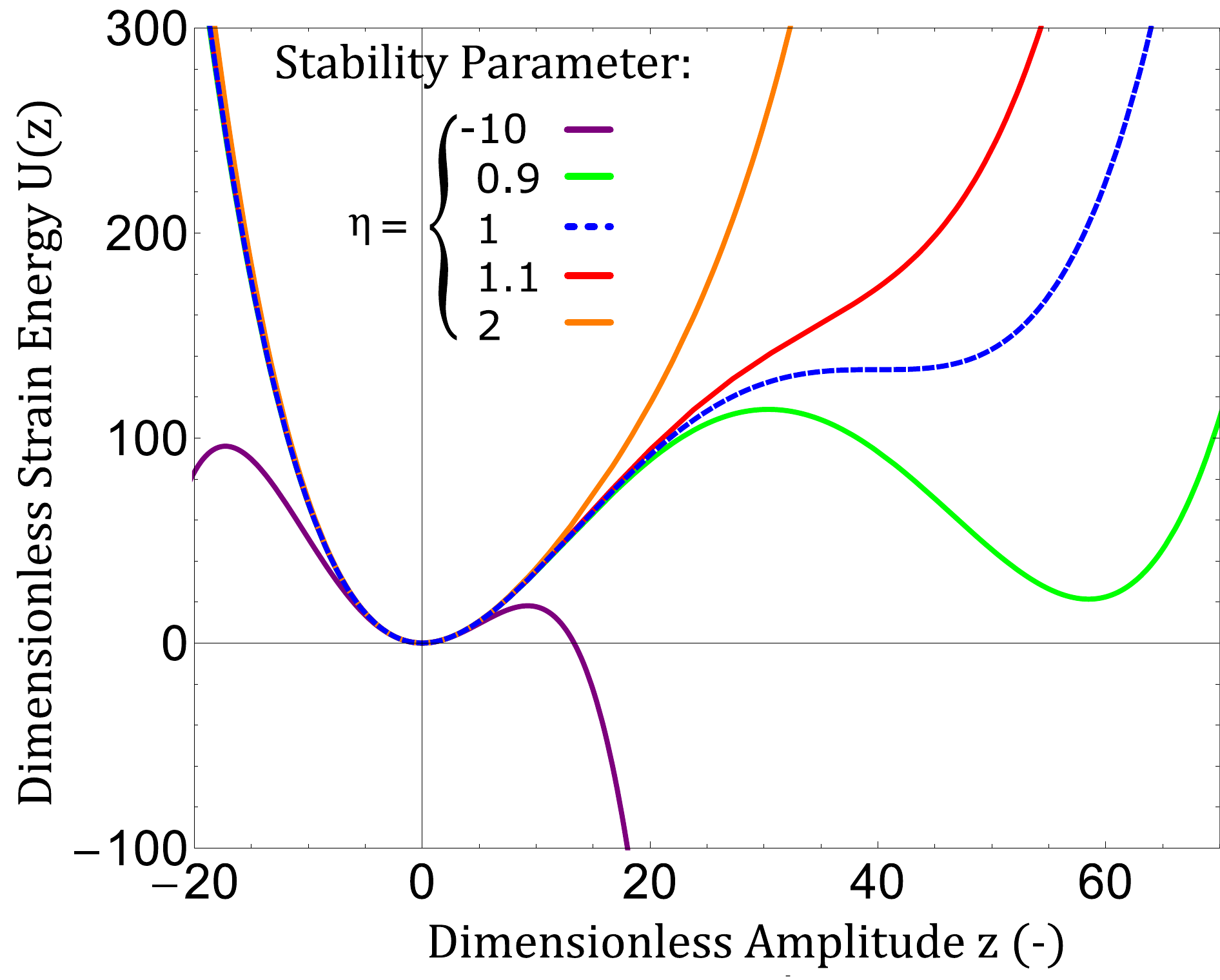}
	\caption{Various functions of the strain energy for different values of the stability parameter \(\eta\) plotted over the dimensionless displacement z.}
	\label{fig:equilibrium}
\end{figure}

\noindent We start our analytical investigations by determining the equilibrium positions of the mechanical harvester system. Due to the asymmetric force term \(\alpha x^2\) the system can take on several equilibrium positions where the restoring force onto the oscillator given in equation \eqref{eq:duff_eom_dimless} vanishes, that is where \(z + \sigma z^2 + \frac{\sigma^2}{4} \eta z^3 = 0\). Equivalently, the equilibrium positions are located at the minima of the energy function
\begin{equation}
\label{eq:energy}
E\left(z\right) = \frac{1}{2}z^2 + \frac{\sigma}{3} z^3 + \frac{\sigma^2}{16}\eta z^4,
\end{equation}
as shown in Fig. \ref{fig:equilibrium}.
Here, we can distinguish between different stability properties: For values of \(\eta \leq 0\), the system can become unstable if the oscillatory motion around the equilibrium position \(z = 0\) reaches a threshold. For \(0 < \eta < 1\), a second equilibrium position appears. Thus, the system can snap from an oscillatory motion around the equilibrium position at \(z = 0\) to an oscillation around another equilibrium position with \(z \neq 0\). \(\eta = 1\) constitutes the threshold between stable and unstable behaviour, since the system can reach a saddle point. \\
For \(\eta \geq 1\) which is equivalent to \(\beta \geq \frac{\alpha^2}{4 \omega_{\mathrm{0}}^2}\), the system will always oscillate around the equilibrium position \(z = 0\). 
Thus, in order to guarantee a stable operation of the VEH without possible issues originating from instabilities or snapping we would need to fulfil the system parameter requirement \(\beta \ge \frac{\alpha^2}{4 \omega_{\mathrm{0}}^2}\). 
However, as we will show below, this requirement leads to strong bi-stabilities which in turn render a stable but also unforeseeable output of the VEH due to the rather random nature of the external vibratory forces. A different strategy, therefore, is to introduce the snapping deflection \(z_{\mathrm{snap}}\) which is the saddle point of \(E \left(z\right)\) in equation \eqref{eq:energy}, defined for \(\eta \le 1\) by 
\begin{equation}
\label{eq:snap}
z_{\mathrm{snap}} = | \frac{2}{\eta \sigma} \left(\sqrt{1-\eta}-1\right)|\overset{\eta \rightarrow 0}{\rightarrow} \frac{1}{|\sigma|}
\end{equation}
and guarantee that the oscillation amplitude never exceeds the snapping deflection for all operation points and times, i.e. \(z\left(\tau\right) < z_{\mathrm{snap}} \forall \: t, \sigma, Q, r, \eta\). Assuming the largest possible oscillation amplitude to be at resonance, we can approximately demand that \(1 > Q |\sigma|\) when \(\eta \approx 0 \). Yet, as we will discuss below, the 1\textsuperscript{st} order parametric resonance is reached only at a critical amplitude of \(z_{\mathrm{crit}} \approx \frac{1}{Q |\sigma|} \). This condition cannot be fulfilled at the same time as the demand \(1 > Q |\sigma| \). Consequently, this strategy would not yield a first order parametric resonance and thus, the demand \(1 > Q |\sigma|\) is dropped.

\begin{figure}[htbp]
	\centering
	\includegraphics[width=0.5\linewidth]{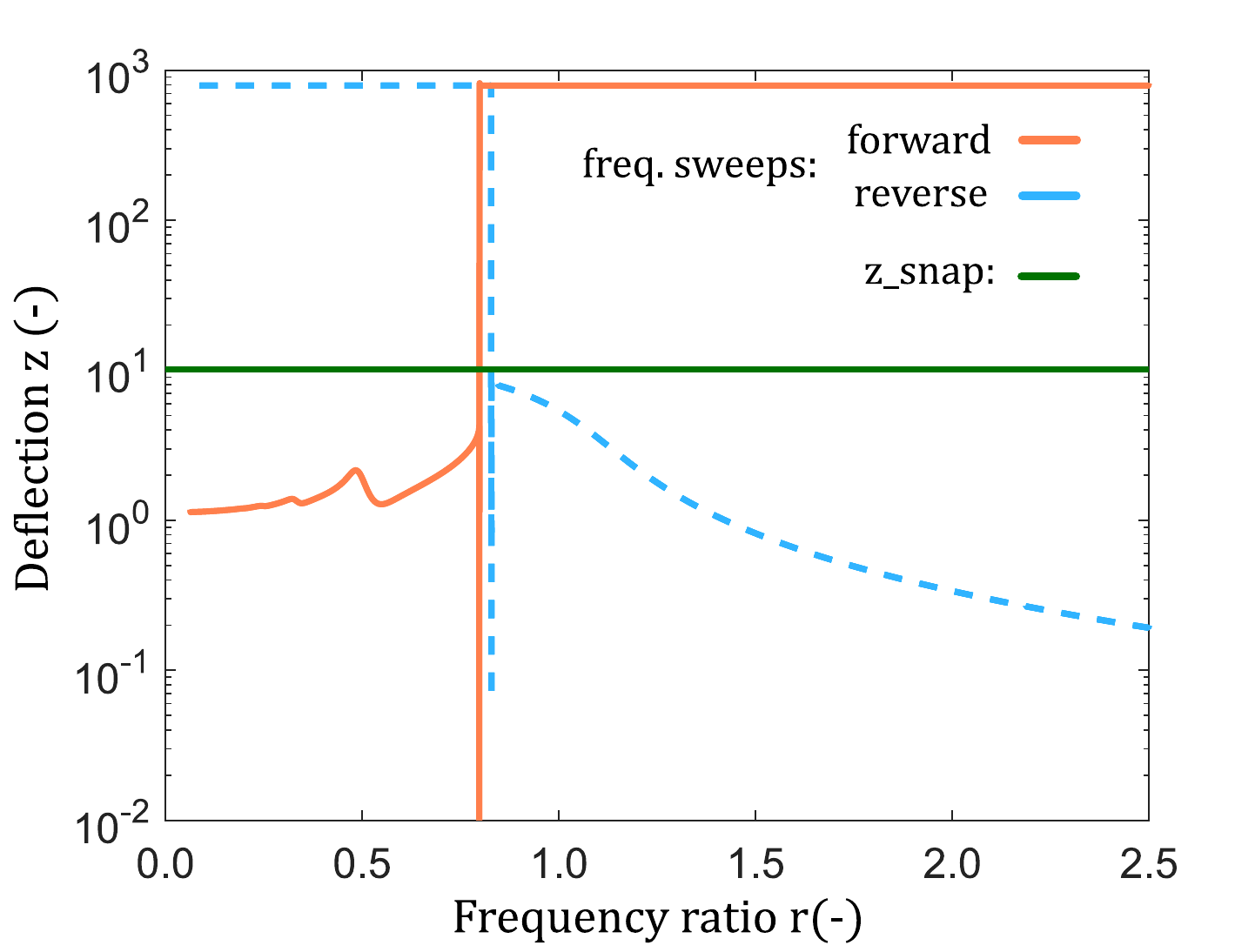}
	\caption{Frequency sweeps of the dimensionless displacement (forward in light red, reverse in light blue) contrasted with the snapping deflection \(z_{\mathrm{snap}}\) (green horizontal line). Parameter set: \(Q = 10, \sigma = 0.1, \eta = 0.05\). }
	\label{fig:snap}
\end{figure}
We depict the snapping behaviour in Fig. \ref{fig:snap}, where the forward and reverse frequency sweeps are shown in light red and light blue, respectively. The snapping deflection calculated in equation \eqref{eq:snap} is given by the green horizontal line. When, for either of the sweep directions, the values of \(z\left(t\right)\) approach the green line, the system snaps to a nonzero equilibrium position (not shown in this figure). In the case of the light red line denoting the forward sweep, a strong transient overshoot takes place, when the snapping deflection is reached: The deflection first goes down to very small values (close, but not equal to zero), before the nonzero equilibrium position at a higher deflection is reached. Since this occurs within a short period of time, spanning only a few oscillations, we attribute this to dynamic effects at the transition point.\\
\subsection{Parametric excitations in the weakly nonlinear Helmholtz-Duffing equation}
\label{parexc}
\noindent In order to understand the relation of our system described by equation \eqref{eq:duff_eom_dimless} to parametric amplification, we define the deflection from parametric amplification, \(z_{\mathrm{pa}}\), and deflection independent of parametric amplification, \(z_0\), and use the ansatz 
\begin{equation}
\label{eq:ansatz}
z := z_{\mathrm{pa}} + z_{0}
\end{equation}
with \(z_0 = A_0 \sin\left(r \tau + \phi_0\right)\), where the amplitude \(A_0\) and phase \(\phi_0\) will be defined below. Next, we insert the ansatz given in equation \eqref{eq:ansatz} into equation \eqref{eq:duff_eom_dimless} and linearise around \(z_0\) for small \(z_{\mathrm{pa}}\) which results in 
\begin{align}
 & z_{\mathrm{pa}}'' + \frac{1}{Q} z_{\mathrm{pa}}' + \left(1 + 2\sigma A_0 \sin\left(r \tau + \phi_0\right) + \frac{3\sigma^2 \eta A_0^2}{8} - \frac{3\sigma^2 \eta A_0^2}{8}\cos\left(2 r \tau + 2 \phi_0\right)\right) z_{\mathrm{pa}} \nonumber \\
= & \frac{\sigma A_0^2}{2} \cos\left(2 r \tau + 2 \phi_0\right) + \frac{\sigma^2 \eta A_0^3}{16} \cos\left(3 r \tau + 3 \phi_0\right). & \label{eq:lin_long}
\end{align}
%
Since only time-dependent terms contribute towards the performance ratio (see equation \ref{eq:performance}), we have neglected static terms in equation \eqref{eq:lin_long} and additionally use the Helmholtz-Duffing oscillator within the first order harmonic balance \cite{Landau1982, Kaajakari2004} given by 
\begin{equation}
\label{eq:harmbal}
z_0'' + \frac{1}{Q} z_0' + \left(1 + \left(\frac{3 \sigma^2}{16}\eta  - \frac{5\sigma^2}{6}\right) A_0^2\right) z_0 = \sin\left(r \tau\right)
\end{equation}
to determine the amplitude \(A_0\) and phase \(\phi_0\) \cite{Landau1982, Duffing1918, Nabholz2018} by solving the nonlinear algebraic steady-state equation
\begin{equation}
\label{eq:hd_amp}
A_0^2 = \left(\left( 1 + \left(\frac{3 \sigma^2 \eta}{16} - \frac{5\sigma^2}{6} \right) A_0^2 - r^2\right) + \left(\frac{r}{Q}\right)^2 \right)^{-1}.
\end{equation}
The phase can be deduced from 
\begin{equation}
\label{eq:hd_phase}
\tan\left(\phi_0\right) = \frac{r}{Q} \left(r^2 - 1 - \left(\frac{3\sigma^2}{16} \eta - \frac{5\sigma^2}{6}\right) A_0^2\right)^{-1}. 
\end{equation}
By close inspection of equation \eqref{eq:lin_long}, we recognize a variant of the damped Mathieu equation \cite{Mathieu1868} for the variable \(z_{\mathrm{pa}}\) with periodic forcing terms on the right hand side. More importantly, on the left hand side, we find frequency or rather stiffness modulation terms at modulation period \(\frac{2\pi}{r}\) with amplitude \(2\sigma A_0\) and at modulation period \(\frac{\pi}{r}\) with amplitude \(b := \frac{3\sigma^2\eta}{8} A_0^2\). Since the overall stiffness modulation term is \(\frac{2\pi}{r}\)-periodic, we apply the Floquet ansatz \cite{Rand2005} \(z_{\mathrm{pa}} = \frac{1}{2} e^{\gamma\tau} F\left(\tau\right)\) with the complex valued \(\frac{2\pi}{r}\)-periodic function \(F\left(\tau\right)\), i.e. \(F\left(\tau\right) = F\left(\tau + \frac{2\pi}{r}\right)\), and the characteristic exponent \(\gamma \in \mathbb{C}\). By defining \(F\left(\tau\right) = \sum_{n=-\infty}^{\infty} f_n e^{i n r \tau}\) with \(f_n \in \mathbb{C}\) and \(i = \sqrt{-1}\) we are able to rewrite equation \eqref{eq:lin_long} by 
\begin{align}
\label{eq:floquet}
& \frac{1}{2} e^{\gamma\tau} \sum_{n = -\infty}^{\infty} \left(\left(c^2 + \frac{1}{Q} c + 1 + b \right)f_n - i\sigma A_0\left(f_{n-1} e^{i\phi_0} - f_{n+1}e^{-i\phi_0}\right) - \frac{b}{2} \left(f_{n-2}e^{2 i \phi_0} + f_{n+2} e^{-2 i \phi_0}\right)\right) e^{i n r \tau} \\ \nonumber
 = & \frac{\sigma A_0^2}{4} \left(e^{i\left(2 r \tau + 2\phi_0 \right)} + e^{-i\left(2 r \tau + 2 \phi_0\right)}\right) + \frac{b}{4} A_0 \left(e^{i\left(3 r \tau + 3 \phi_0\right)} + e^{-i\left(3 r \tau + 3 \phi_0\right)}\right) 
\end{align}
with the short notation \(c := \gamma + i n r\).\\
According to Floquet theory \cite{Rand2005}, the stationary state solutions of \(z_{\mathrm{pa}}\) can be either \(\frac{2\pi}{r}\)- or \(\frac{4\pi}{r}\)-periodic. Thus, without loss of generality, the imaginary part of the characteristic exponent is given by \(\operatorname{Im}\{\gamma\} = 0\) or by \(\operatorname{Im}\{\gamma\} = \frac{r}{2}\) for the \(\frac{2\pi}{r}\)- or \(\frac{4\pi}{r}\)-periodic solutions, respectively. For the case of stable oscillatory steady-state solution we also need to consider \(\operatorname{Re}\{\gamma\} = 0\).\\
\subsection{Critical points of parametric resonances}
\noindent The Strutt diagram \cite{Strutt1883, Rayleigh1887}, the well-known stability diagram of the Mathieu equation, features the transition from stable steady-state solutions to regions of instability where periodic solutions grow exponentially in time. These so-called parametric resonances (in the following denoted by the index \(PR\)) will occur whenever the system parameters enter the instability regions across the critical points \cite{Minorsky1962}. In the concrete case of equation \eqref{eq:lin_long}, we will encounter parametric resonances oscillating at frequencies \(f_{\mathrm{PR}} = \frac{n r}{2}, n \in \mathbb{N}\), for the n-th order parametric resonance. As we will show below, the critical points of the n-th order parametric resonances are well within reach of the system parameters, if \(r \approx \frac{2}{n}\), thus \(f_{\mathrm{PR}} \approx 1 \: \forall \: n \in \mathbb{N}\). \\
In order to find the critical points we solve for the homogenous version of equation \eqref{eq:floquet}, the is \(\bm{H} \bm{f} = 0\) with the infinite solution vector \(\bm{F} = \{...,f_{-1}, f_0, f_1, f_2, ...\}\) and the infinite Hill matrix \(\bm{H}\) \cite{Brand2015}. With the Kronecker delta \(\delta_{n,m}\), the elements of the Hill matrix read
\begin{align}
\label{eq:hill}
\bm{H}_{nm} = & \left(c^2 + \frac{1}{Q} c + 1 + b\right) \delta_{n,m} + i\sigma A_0 \left(e^{-i\phi_0} \delta_{n+1,m} - e^{i\phi_0} \delta_{n-1,m}\right) \\ \nonumber
& - \frac{b}{2} \left(\delta_{n+2,m} e^{-2 i \phi_0} + \delta_{n-2,m} e^{2 i \phi_0}\right). 
\end{align}
Note that the trivial solution \(f_n = 0 \: \forall \: n \) solves the homogeneous Mathieu equation \(\bm{H} \bm{f} = 0\) for the steady-state. However, this trivial solution becomes unstable at the critical points, in favour of a non-trivial solution which can only exist if \(\mathrm{det}\left(\bm{H}\right)= 0\). The line in parameter space, where \(\operatorname{Re}\{\gamma\} = 0\) marks the critical points or rather the transition from stable to unstable solutions of \(\bm{H} \bm{f} = 0\). Therefore, we will numerically find the line defined by \(\mathrm{det}\left(\bm{H}\right)= 0\) in the parameter space of \(A_0\) and \(r\) for given values of the system parameters \(Q, \sigma, \eta\) for \(\gamma = i\frac{r}{2}\) and \(\gamma = 0\) to determine the critical points of the 1\textsuperscript{st}, 3\textsuperscript{rd}, 5\textsuperscript{th},... and 2\textsuperscript{nd}, 4\textsuperscript{th}, 6\textsuperscript{th},... parametric resonances, respectively. We remark that \(\mathrm{det}\left(\bm{H}\right)\) does not depend on the phase factor \(\phi_0\). Moreover, we ensure the convergence of the results with the inevitable numerical truncation of the Hill matrix. \\
We calculate the critical power required for the onset of the n-th parametric resonance according to \(p_{\mathrm{crit}} = \frac{1}{Q} A_0^{cr^2}r_{cr}^2\) where the critical points \(\{A_0^{\mathrm{cr}}, r_{\mathrm{cr}}\}\) are found numerically by solving for \(\mathrm{det}\left(\bm{H}\right) = 0\). 
\begin{figure}
	\centering
	\includegraphics[width=0.6\linewidth]{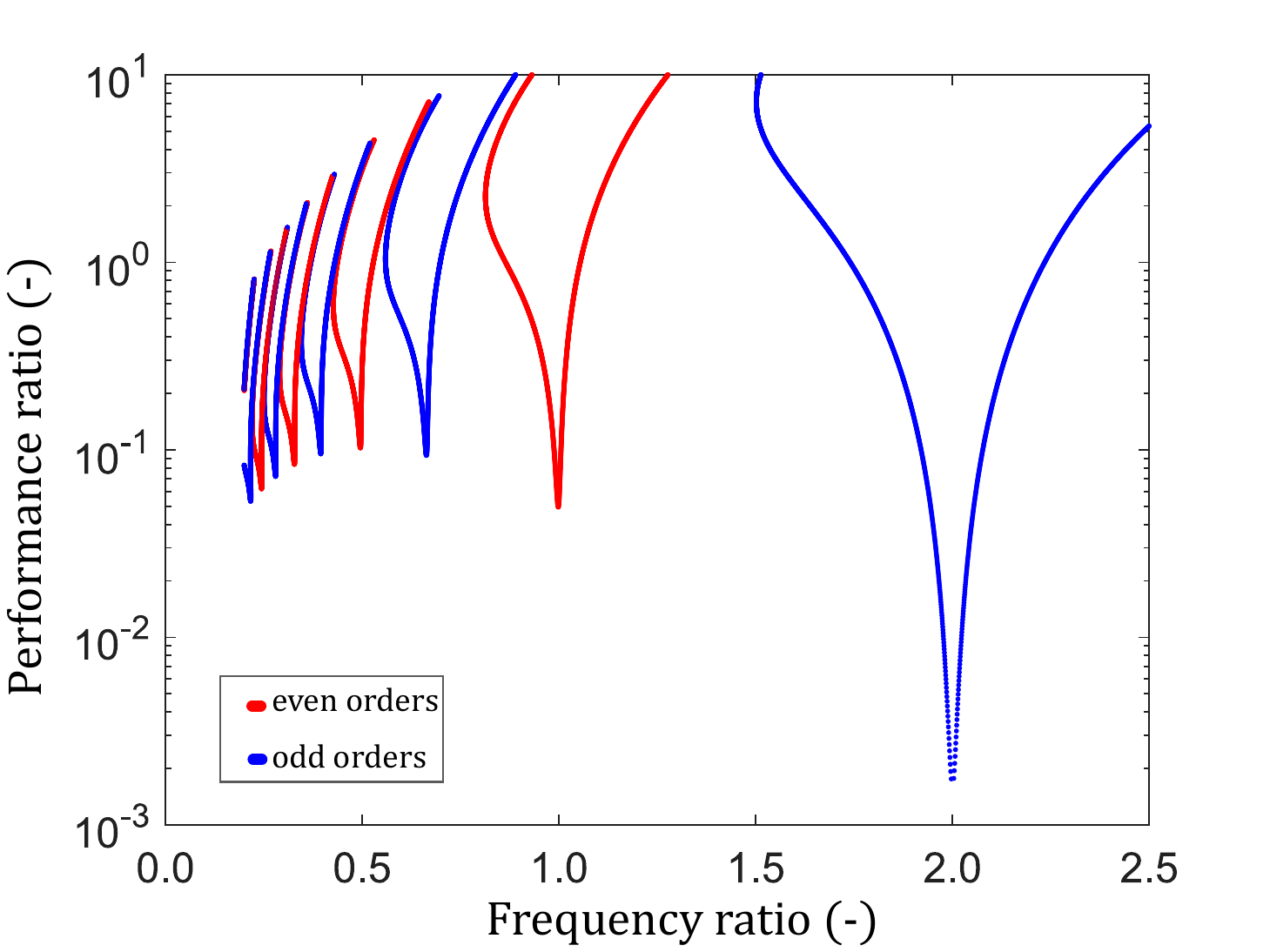}
	\caption{Example plot of the critical point curves for the parameter set \(Q = 100, \sigma = 0.05, \eta = 1.01\). Odd orders of parametric resonance in blue, even orders in red.}
	\label{fig:critical_temp}
\end{figure}
\noindent The results for an example parameter set \(Q = 100, \sigma = 0.05, \eta = 1.01\) are shown in Fig. \ref{fig:critical_temp}, with the odd orders of of parametric resonance depicted in blue and the even orders in red.\\
For the first order parametric resonance, an exact 1\textsuperscript{st} order harmonic balance solution exists \cite{Rand2005}. It is given by 
\begin{equation}
A_{0, \mathrm{PR}}^{1\textsuperscript{st}} = \frac{1}{|\sigma|}\left(\left(2-r\right)^2 + \frac{1}{Q^2}\right)^{0.5} \label{eq:crit}
\end{equation}
%
%
\subsection{Parametric amplification below the critical point}
\label{subcrit}
\noindent Apart from parametric resonances appearing above the critical points, equation \eqref{eq:lin_long} also provides finite oscillatory solutions even below the critical points (with the index \(BT\) denoting parameters below the critical point) as a result of the forcing terms on the right hand side. However, due to the \(2\pi\)-periodicity of the forcing terms, the \(4\pi\)-periodic solutions, that is the 1\textsuperscript{st}, 3\textsuperscript{rd}, 5\textsuperscript{th},... order parametric excitations will remain zero below the critical points. This insight becomes clear from coefficient comparison of the exponential functions \(e^{i...\tau}\) in equation \eqref{eq:floquet}. For the 2\textsuperscript{nd}, 4\textsuperscript{th}, 6\textsuperscript{th},... parametric excitations, on the other hand, the trivial solution does not solve equation \eqref{eq:lin_long} or rather equation \eqref{eq:floquet}, and therefore, parametric amplification effects are expected. Overall, since we are interested in the steady-state solution of the \(2\pi\)-periodic solutions, we will solve equation \eqref{eq:floquet} for \(\operatorname{Im}\{\gamma\} = 0\) and \(\operatorname{Re}\{\gamma\} = 0 \) in the form
\begin{equation}
\label{eq:subcrit}
\bm{f}_{\mathrm{BT}} = \bm{H}^{-1} \bm{g},
\end{equation}
with the acceleration vector
\begin{equation}
\left(\bm{G}\right)_n = \frac{\sigma A_0^2}{2} \delta_{n,2} + \frac{\sigma^2 \eta A_0^3}{16} \delta_{n,3},
\end{equation}
for the solution vector \(\bm{f}_{\mathrm{BT}} = \{..., f_{-1}^{\mathrm{BT}}, f_{0}^{\mathrm{BT}}, f_{1}^{\mathrm{BT}}, f_{2}^{\mathrm{BT}}, ...\}\). In the following, we refer to the solution of equation \eqref{eq:subcrit} as the sub-critical solution given by \(z_{\mathrm{pa}}^{\mathrm{BT}} = \sum_{n=0}^{\infty} |f_n^{\mathrm{BT}}| \cos\left(n r \tau + \theta_n\right)\) with \(\theta_n = \frac{\operatorname{Im}\{f_n^{\mathrm{BT}}\}}{\operatorname{Re}\{f_n^{\mathrm{BT}}\}}\). Recall that the Hill matrix will always be invertible below the critical point for a given parameter set \(\{\sigma, \eta, Q, A_0, \phi_0\}\). Note, however, that the solution of the Helmholtz-Duffing equation, see equation \eqref{eq:harmbal}, shows a hysteresis for the steady-state quantities \(A_0, \phi_0\) for certain parameter sets and therefore, we also calculate \(z_{\mathrm{pa}}^{\mathrm{BT}}\) for a forward and a reverse sweep of the frequency ratio \(r\).
\section{Results}
\label{simulation}
\noindent We investigate the effects of parametric amplification in order to estimate the parameter range needed for a parametrically amplified VEH. Our optimization goal is increased bandwidth, not peak power, since this increases the frequency-averaged volumetric power density of a harvester array by lowering the number of substructures necessary \cite{Lamprecht2019}. This poses a challenge towards the design of harvester structures \cite{Harne2013}. \\
The acceleration ratio given in equation \eqref{eq:sigma} shows that \(\sigma \propto \alpha\), and nonlinear behaviour of the structure (and thus, parametric resonance) can be achieved by increasing the absolute value of \(\sigma\). With the dimensionless representation given in equation \eqref{eq:duff_eom_dimless}, only three system parameters need to be investigated: the quality factor \(Q\), the acceleration ratio \(\sigma\) and the stability parameter \(\eta\). \\
Our results combine the information four models, or rather simulation strategies: \\
Firstly, our transient simulation of the Helmholtz-Duffing equation establishes the dynamic behaviour we can expect from the oscillator system. The transient simulation is carried out by simulating forward and reverse sweeps of equation \eqref{eq:duff_eom_dimless}, with the frequency ratio given in equation \ref{eq:ratio_r} as the sweep parameter, representing a linear change of the excitation frequency \(\omega_{\mathrm{exc}}\). \\
The second model provides the baseline: We employ the steady-state solution of the first order harmonic balance for the Helmholtz-Duffing oscillator, that is the solution of equation \eqref{eq:hd_amp}. In the case of multiple solution branches due a shift of the resonance peak over frequency, we assume the stability properties of the basic Duffing oscillator to hold true \cite{Duffing1918}. \\
The last two models are both the result of our analytical derivations: \\
The sub-critical model provides us with an approximate solution of the system behaviour below the critical point. It diverges from the baseline of the harmonic balance approach, whenever parametric amplification occurs and thus it should be in agreement with our transient simulation. Above the critical point, the sub-critical solution has no physical relevance and is thus omitted from the representation.\\
Lastly, the critical point model indicates the onset of parametric resonance where it intersects with the harmonic balance solution. It is important to note that the critical point model does not model the actual deflection of the system. \\
Combining the information from our four models yields the basis for deciding on the validity of our analytic approach, but most importantly it provides a quantitative assessment of the system parameters in order to arrive at a parameter set that provides the best harvesting capability for our broadband approach.\\
\begin{figure}[htbp]
	\centering
	\begin{minipage}{1.0\textwidth}
		\subfloat[Parameter set: \(Q = 100, \sigma = 10^{-4}, \eta = 1.01\)]
		{
			\includegraphics[width=1.0\textwidth]{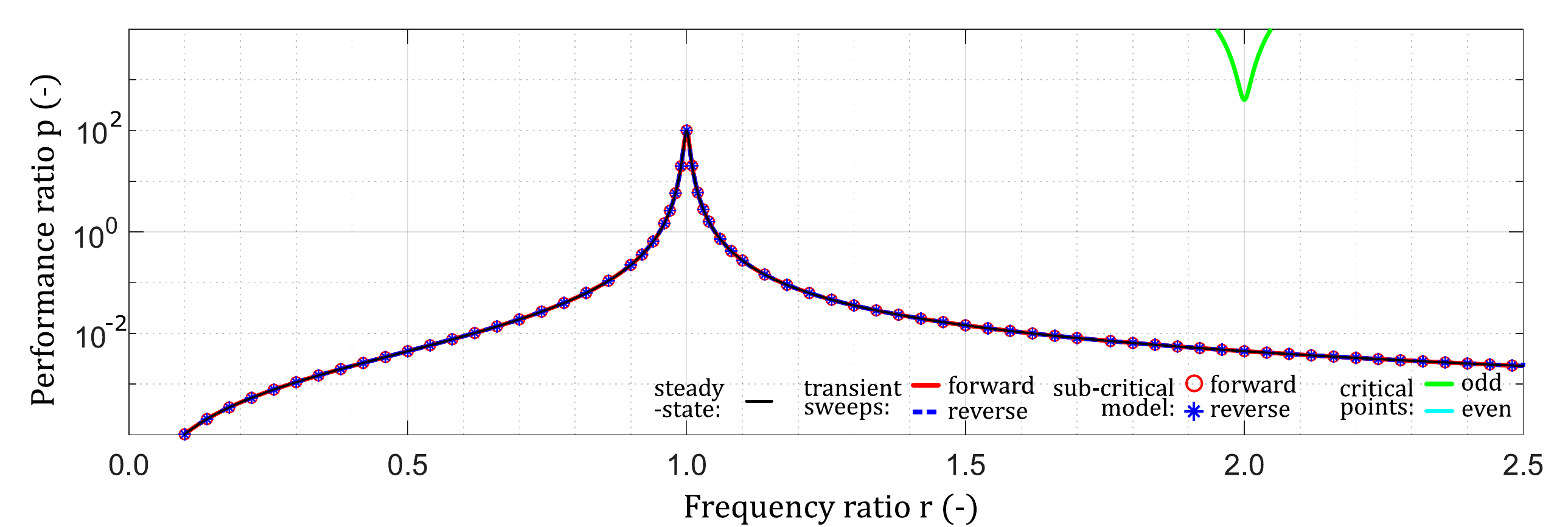}
			\label{SUBFIGURE:test1}
		}
		\hfil
		\subfloat[Parameter set: \(Q = 10, \sigma = 0.1, \eta = 0.05\)]
		{
			\includegraphics[width=1.0\textwidth]{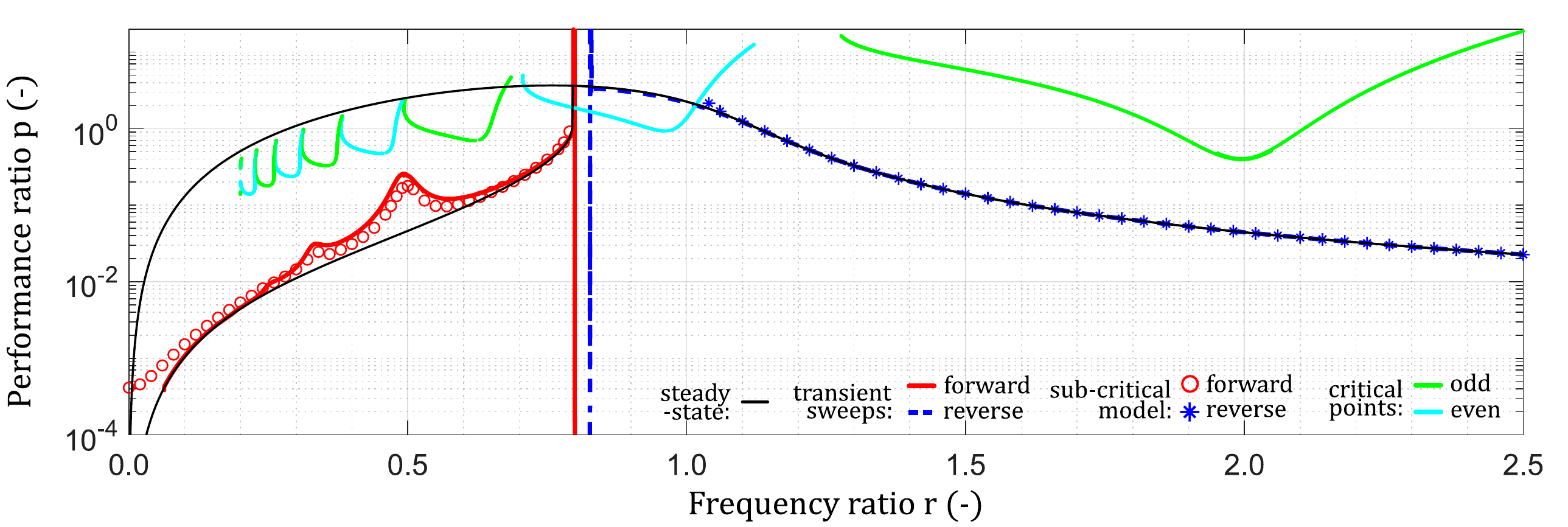}
			\label{SUBFIGURE:test2}
		}
		\hfil
		\subfloat[Parameter set: \(Q = 10, \sigma = 0.01, \eta = 10^4\)]
		{
			\includegraphics[width=1.0\textwidth]{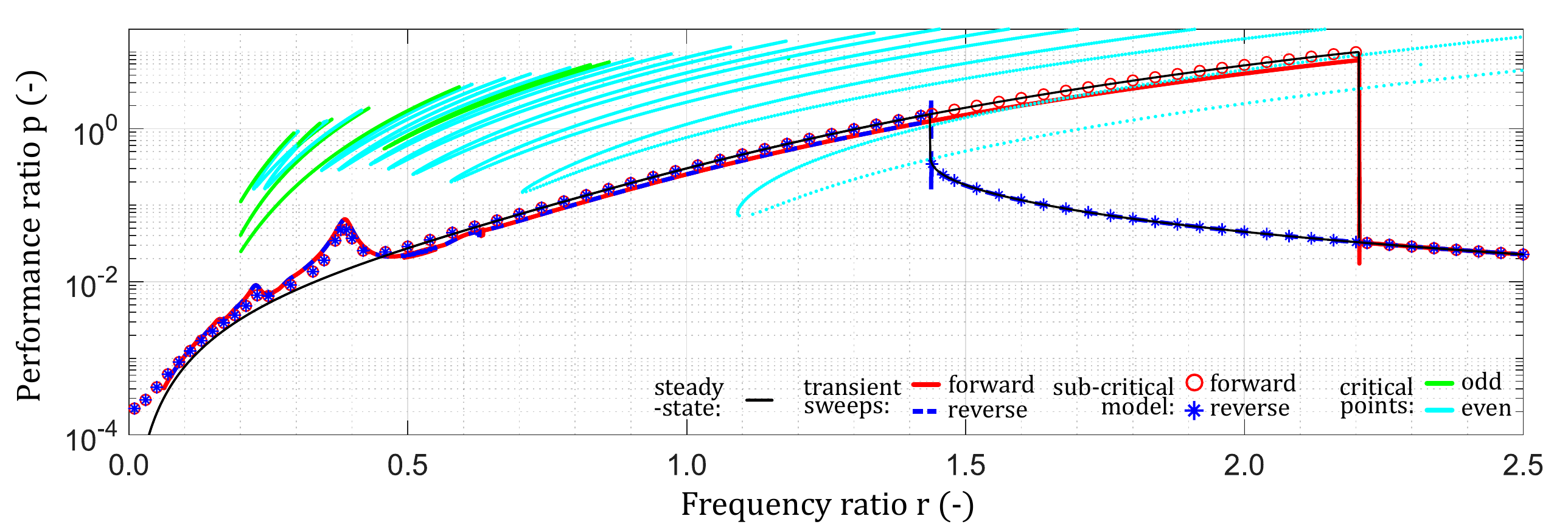}
			\label{SUBFIGURE:test3}
		}
	\end{minipage}
	\caption{Transient simulations of equation \eqref{eq:duff_eom_dimless} in forward (red) and reverse (blue, dashed) direction. The first order harmonic balance solutions from equation \eqref{eq:hd_amp} are shown for forward (black, solid) and reverse sweep (black, dashed). The solutions of the sub-critical model obtained from equation \eqref{eq:subcrit} are shown with red circles for the forward sweep and blue stars for the reverse sweep.  The critical points for even (cyan) and odd (green) orders of parametric resonance are obtained from equation \eqref{eq:hill}, setting \(\mathrm{det}\left(\bm{H}\right) = 0\).}
	\label{fig:results}
\end{figure}
\noindent Comparing the five plots in Fig. \ref{fig:results} and \ref{fig:results2} showcases the influence of the parameters \(Q, \sigma, \eta\) on the performance ratio defined in equation \eqref{eq:performance} in the parametrically amplified regime. The horizontal axis shows the dimensionless frequency ratio \(r\) and the vertical axis shows the dimensionless performance measure \(p\) as defined in equation \eqref{eq:performance}. Transient simulations of equation \eqref{eq:duff_eom_dimless} are shown using a frequency sweep in forward (red) and reverse (blue, dashed) direction. For comparison, the first order harmonic balance solutions for the Helmholtz Duffing oscillator, given in equation \eqref{eq:hd_amp} for forward (black, solid) and reverse sweep (black, dashed) are given. The solutions of the sub-critical model obtained from equation \eqref{eq:subcrit} are shown with red circles for the forward sweep and blue stars for the reverse sweep. Note that the solutions of the transient simulation only become visible, when they diverge from the sub-critical model. Lastly, the critical points for even (cyan) and odd (green) orders of parametric resonance are shown as obtained from setting the determinant of the Hill matrix in equation \eqref{eq:hill} to zero, compare Fig. \ref{fig:critical_temp}. We only calculate the critical points down to a frequency ratio of \(r = 0.2\), since the curves move further together for smaller values and obscure rather than enhance the representation. The sub-critical model is only shown as long as no critical point curve intersects the first order harmonic balance solution, since this marks the onset of parametric resonance for which \(z_{\mathrm{pa}}\) gains amplitude and our linearisation in equation \eqref{eq:lin_long} no longer holds true.\\
Fig. \ref{SUBFIGURE:test1} provides the starting point of our analysis with the parameter set \(Q = 100, \sigma = 10^{-4}, \eta = 1.01\), where no parametric resonance is expected from the transient simulation: The results of forward and reverse transient simulation are both identical to the analytic steady-state solution for the driven, damped harmonic oscillator with \(\sigma = 0\). The green curve shows the critical point of first order parametric resonance, the critical points of higher order parametric resonance are not within the plot range. No parametric resonance is predicted from the transient simulation and none occurs. This is confirmed by the sub-critical solution that predicts exactly the same values as the transient simulation (thus partially obscuring their plotted lines).\\
Following up on Fig. \ref{fig:snap}, simulations for the same parameter set \(Q = 10, \sigma = 0.1, \eta = 0.05\) are shown in Fig \ref{SUBFIGURE:test2}. It becomes apparent how small values of \(\eta\) lead to snapping effects which change the centre of oscillation, indicated by the two minima of the strain energy plot for  \(\eta = 0.9\) in Fig. \ref{fig:equilibrium}. In contrast to Fig. \ref{fig:snap}, we have limited the plot range in vertical direction. The transition from one equilibrium position to another poses the risk of mechanical failure. Thus, such a parameter set is not useful for harvesting applications and we need to ensure the parameter limit \(\eta > 1\). Yet, Fig. \ref{fig:snap} provides a good validation example for our model and the underlying assumptions: The sub-critical model follows the transient simulation for forward and reverse sweep up until the critical point curve is crossed from both directions around \(r \approx 0.8\) and \(r\approx 1.0 \), respectively. The only considerable divergence occurs for small values \(r < 0.1\) at the very left end of the curve. Here, the transient simulation lies below the sub-critical model. Due to minor importance of such small frequencies, we have excluded this area from our detailed analysis and also limited the plot range of the critical point range to \(r>0.2\), both for reasons of clear representation and relevance. \\
Fig. \ref{SUBFIGURE:test3} depicts the limits of the linearisation in equation \ref{eq:lin_long} with the parameter set \(Q = 10, \sigma = 0.01, \eta = 10^4\): For such large values of \(\eta\), parametric amplification leads to bistable Duffing-like behaviour. Furthermore, bistable behaviour, where forward and reverse sweeps yield vastly different values of \(p\), occurs in large ranges of \(r\). Thus, considering the stochastic nature of the vibrations in an industrial environment, we could not be sure that the system response stays on the upper solution branch due to the insufficient knowledge of the initial state. We will aim at a parameter set that minimizes the range of bistable behaviour in order to arrive at a reliable estimate of the performance ratio \(p\) of our VEH design. In total, this parameter set yields only small advantages over a harvester without parametric amplification, since the performance ratio is only slightly increased for smaller frequency ratios \(r\). \\
\begin{figure}[htbp]
	\centering
	\begin{minipage}{1.0\textwidth}
		\subfloat[Parameter set: \(Q = 100, \sigma = 0.05, \eta = 1.01\)]
		{
			\includegraphics[width=1.0\textwidth]{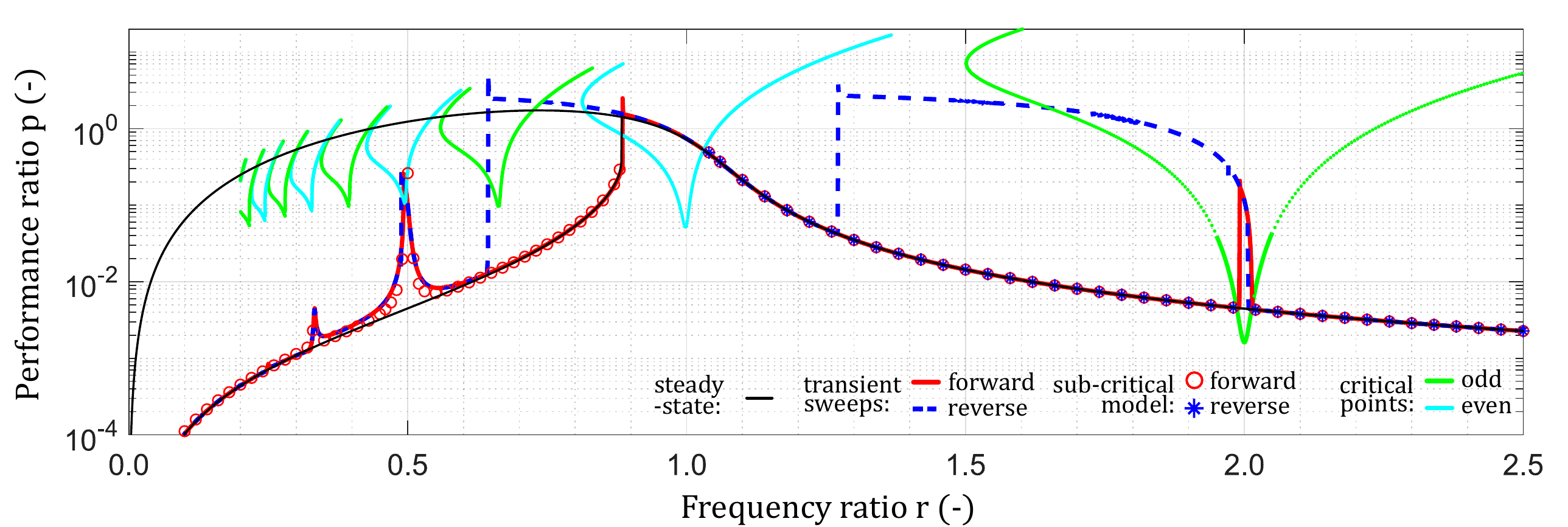}
			\label{SUBFIGURE:test4}
		}
		\hfil
		\subfloat[Parameter set: \(Q = 10, \sigma = 0.5, \eta = 1.01\)]
		{
			\includegraphics[width=1.0\textwidth]{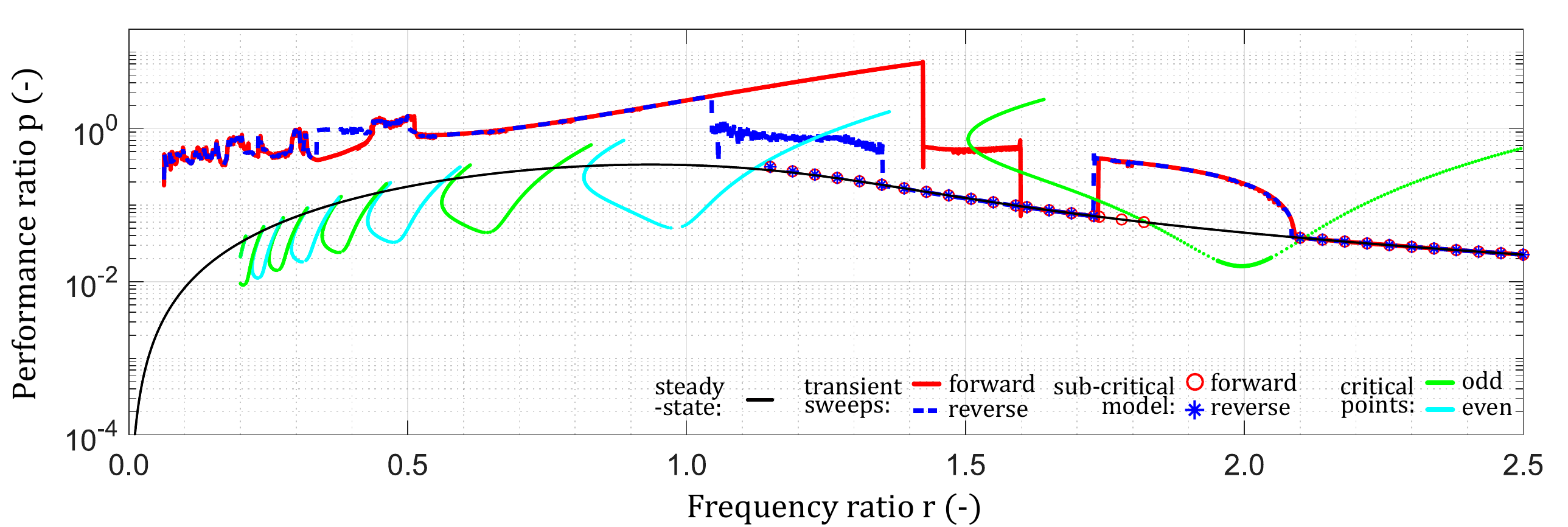}
			\label{SUBFIGURE:test5}
		}
		
	\end{minipage}
	\caption{Transient simulations of equation \eqref{eq:duff_eom_dimless} in forward (red) and reverse (blue, dashed) direction. The first order harmonic balance solutions from equation \eqref{eq:hd_amp} are shown for forward (black, solid) and reverse sweep (black, dashed). The solutions of the sub-critical model taken from equation \eqref{eq:subcrit} are shown with red circles for the forward sweep and blue stars for the reverse sweep.  The critical points for even (cyan) and odd (green) orders of parametric resonance are obtained from equation \eqref{eq:hill}, setting \(\mathrm{det}\left(\bm{H}\right) = 0\).}
	\label{fig:results2}
\end{figure}
\noindent Fig. \ref{fig:results2} focuses on parameters sets for possible implementations. We focus on values of \(\eta \approx 1\), where snapping cannot occur but the frequency shift of the resonance peaks towards higher frequencies is still limited.\\
Fig. \ref{SUBFIGURE:test4} with the parameter set \(Q = 100, \sigma = 0.05, \eta = 1.01\) depicts a case where parametric resonance occurs for a quality factor which roughly constitutes the upper limit of what can be expected for our VEH. Here, several orders of parametric amplification occur below the critical point. The sub-critical model provides a very good estimate of the transient system behaviour. Note that the first order parametric resonance (on the right hand side, where the green curve intersects) marks exactly the onset of parametric amplification in both sweep direction. Here, forward and reverse sweeps yield different results, i.e. hysteresis occurs, entailing a large area of bistable behaviour on the right hand side. Here, we can only predict the onset of parametric amplification given by the intersection of critical point model and sub-critical or steady-state model, but we cannot predict, where the higher deflection solution branch will be left in favour of the lower one. Similar response curves are proposed for broadband energy harvesting applications, e.g. in \cite{Harne2013} where the best-case scenario is always assumed. This poses a problem, since in the case of stochastic vibrations, there exists no reliable means for predicting which solution branch the system will assume \cite{Daqaq2009}. As Fig. \ref{SUBFIGURE:test4} clearly shows, the frequency range of high performance ratios is much wider for reverse than for forward frequency sweeps, due to jump phenomena. For this reason, we aim to minimise the hysteresis effects. We note that focussing on values of \(\eta \approx 1\) is not sufficient to avoid different performance ratios for forward and reverse sweep. Thus, we need to further increase the asymmetric nonlinearity by increasing \(\sigma\). We also expect better broadband performance for lower quality factors, since lowering the quality factor also broadens the peaks of the response curve. 
For the parameter set in Fig. \ref{SUBFIGURE:test5}, \(\eta\) was tuned to achieve broadband amplification. Due to complex system dynamics, many order of parametric resonance occur and the effects overlap, leading to a raised performance ratio for \(r < 1.3\). It should be noted that even higher values of \(\sigma\) lead to a deviation of the modelled onset of first order parametric resonance from the prediction of the critical point model. From the transient simulation, we obtain large regions with a raised performance ratio, even though the overall level is lower than the resonant peak. Furthermore, the areas of bistable behaviour are small, thus making the harvested power level more predictable and independent of the machine-specific (and thus, application-specific) vibration characteristics. Thus, we arrive at a recommendation of this (or a similar) parameter set for future design implementations of our model for a broadband VEH. \\
Whilst Fig. \ref{SUBFIGURE:test1} - \ref{SUBFIGURE:test4} provide validation of our models and thus justification for relying on the predictions of our combined simulations for the final device layout, they also aided in arriving at the parameter set simulated in Fig. \ref{SUBFIGURE:test5} that enables broadband harvesting with high performance ratios in both sweep directions.
\section{Conclusion}
\label{conclusion}
\noindent From structural mechanic, we derived nonlinear stress-strain relations for tuning the nonlinear properties of the oscillatory system and introduced the concept of employing magnetic or electric fields as supporting sources of vibrational energy using field-induced striction. These fields occur naturally in industrial environments, but are too weak to be harvested using conventional techniques. \\
To assess the parameter requirements for a parametrically amplified VEH, we formulated a single mode approach based on the Helmholtz-Duffing oscillator and employed a combination of four different simulation strategies: Transient simulations of the parametrically amplified Helmholtz-Duffing equation, an analytical sub-critical model, a critical point model for even and odd orders of parametric resonance and a baseline steady-state first order harmonic balance model. A comparison of analytical and transient system models shows that parametric amplification phenomena can be traced back to a combination of critical and sub-critical parametric amplification.\\
Consequently, driving a VEH model in the regime of parametric amplification enables us to increase the range of covered frequencies for energy harvesting, see especially Fig. \ref{SUBFIGURE:test5}. As a result, fewer individual oscillator structures are needed to achieve broadband harvesting compared to regular harvester array configurations with linear oscillators. Furthermore, our approach ensures a response function with increased achievable broadband power for forward and reverse sweeps which constitutes an improvement over classical nonlinear energy harvesting approaches. Thus, we developed a comprehensive model for parametrically amplified harvesting of the ambient energy of magnetic or electric fields in industrial environments as a means of achieving maintenance-free energy supply to a wireless sensor node.\\
Even though the focus of this work was the derivation and implementation of the nonlinear model and no design has been fabricated yet, our analysis takes into account the boundary conditions and limitations imposed by realistic operating conditions. Thus our theoretical assessment of the feasibility of such a VEH design provides the basis for the design of future test devices (see Section \ref{outlook}).\\ 
The main limitation of our analytical critical point model lies with this unpredictability of bistable nonlinear states. Since \(\sigma\) is  proportional to the external acceleration amplitude, its real value fluctuates along with the external acceleration that is harvested. Thus, a clear prediction of the system behaviour as suggested in Fig. \ref{fig:results} and \ref{fig:results2} is always subject to some uncertainty pertaining to the exact operating conditions and should be quantified using a test design. A further limitation is the fact that behaviour above the critical point can only be predicted quantitatively by our transient simulation, since for large values of the deflection due to parametric amplification, \(z_{\mathrm{pa}}\), our linearisation in equation \eqref{eq:lin_long} no longer holds true.  \\
\section{Outlook}
\label{outlook}
\noindent As our assessment of the limitations of our model in Section \ref{conclusion} shows, our theoretical work should be experimentally verified using a device based on the basic design approach in Section \ref{design}. Such a device needs to combine vibrational excitation with field-induced striction, where field-dependent stress-strain curves of magneto- or electro-strictive materials can be used to achieve nonlinear behaviour, see \cite{Datta2010}, Fig. 4. In order to design a device with the parameter set for maximum exploitation of parametric excitation shown in Fig. \ref{SUBFIGURE:test5}, a quantitative estimation of the strictive properties is necessary. To achieve the lowered quality factors we have proposed, dissipative dampers in the form of viscous clamping material provide a staring point for parameter tuning. 
%
%
\ifCLASSOPTIONcaptionsoff
  \newpage
\fi

\bibliographystyle{IEEEtran}
\bibliography{IEEEabrv,basic_bib}

\end{document}